\documentstyle[12pt,epsf]{article}                                                   
\newcommand{\tr}{\mbox{Tr}}

   \let\m=\mu                      
\let\n=\nu   \let\r=\rho \let\s=\sigma                      
                                                      
  \let\PH=\Phi

\def\0{\over } \def\1{\vec }     \def\2{{1\over2}} \def\4{{1\over4}}            
\def\5{\bar }  \def\6{\partial } \def\7#1{{#1}\llap{/}}                         
\def\8#1{{\textstyle{#1}}}       \def\9#1{{\bf {#1}}}                           
 \def\llp{\hbox to 0pt{\hss /\hskip1.5pt}}    
\def\llo{\hbox to 0.2pt{\hss /}} \def\llq{\hbox to 0pt{\hss /\hskip0.5pt}}      
\def\so{\supset\hbox to 0pt{\hss $\displaystyle -$\hskip1pt}}

\def\<{\langle } \def\>{\rangle }

   \let\hc=\dagger

\let\nn=\nonumber                                                               
\def\bea{\begin{eqnarray}} \def\eea{\end{eqnarray}}                             
\def\beann{\begin{eqnarray*}} \def\eeann{\end{eqnarray*}}                       
\def\beq{\begin{equation}} \def\eeq{\end{equation}}                             
                                                                                
\date{}
\title{
{\large\rm DESY 96-241}\hfill{\large\rm December 1996}\\
{\large\rm HD-THEP-96-50}\hfill\vspace*{3.0cm}\\
Magnetic Screening in the High Temperature Phase\\
of the Standard Model}
\author{W. Buchm\"{u}ller\\ 
{\normalsize\it Deutsches Elektronen-Synchrotron DESY, 22603 Hamburg,
Germany}\\
\vspace*{0.25cm}\\
O. Philipsen\\
{\normalsize\it Institut f\"ur Theoretische Physik, Universit\"at
Heidelberg, 69120 Heidelberg, Germany}\\
\vspace*{2.0cm}\\                     
}                                                                             
\begin{document}                                                                

\setlength{\baselineskip}{18pt}                                     
\maketitle  
\begin{abstract}
\thispagestyle{empty}
\noindent
Non-perturbative effects in the high-temperature phase of the electroweak
theory are characterized by a magnetic screening length. Its size
influences the range of validity of perturbation theory, and it also
determines the critical Higgs boson mass where the first-order phase
transition changes to a crossover. We propose a gauge-invariant definition
of the magnetic screening length and discuss its role in several 
gauge-dependent and gauge-invariant correlation functions.
\end{abstract} 
\setcounter{page}{0}

\newpage

The electroweak phase transition \cite{kirzh} is of great cosmological 
significance because baryon number and lepton number violating processes 
come into thermal 
equilibrium as the temperature approaches the critical temperature of 
the transition \cite{kuzmin}. In recent years the thermodynamics of the
phase transition has been studied in detail by means of perturbation theory
and numerical lattice simulations. As a first step towards the treatment of 
the full standard model, one usually studies the pure SU(2) Higgs model
neglecting the effects of fermions and the mixing between photon and neutral 
vector boson, which can be included perturbatively. We now know that 
the transition is first-order for Higgs boson masses below 70 GeV, and 
that around $\sim 80$ GeV the first-order transition changes to a 
crossover\footnote{For recent reviews, see \cite{jan}.}.

The electroweak transition is influenced by non-perturbative effects
whose size is governed by a `magnetic screening length', the inverse of a
`magnetic mass'. In perturbation theory a magnetic mass appears as a cutoff 
which regularizes infrared divergencies \cite{linde}. 
The size of this cutoff is closely related to the confinement
scale of the effective three-dimensional theory which describes the
high-temperature limit of the SU(2) Higgs model \cite{jak}. 

In a previous paper \cite{buphi} we have determined a gauge-independent
`magnetic mass' from the exponential fall-off of the gauge boson propagator
by means of gap equations. Contrary to perturbation theory, a mass gap
was found in the symmetric phase. A direct consequence was the prediction 
that the first-order phase transition should turn into a crossover at a 
critical Higgs mass below 100 GeV. 
Recently, such a crossover behaviour has indeed been observed  
in numerical lattice simulations for 
Higgs masses larger than about 80 GeV \cite{lai,ran}.
However, contrary to the expectation in \cite{buphi}, the `magnetic mass' 
was not seen in the correlation functions of gauge-invariant
operators which have been studied in detail in numerical simulations
\cite{mon,kajla,tep,neu,gue} and which yield much larger masses. A
`magnetic mass' was only  seen in the numerical study of the gauge boson
propagator in a fixed (Landau) gauge \cite{neu}. 
In the following we shall make some conjectures which may help to resolve this
puzzle and to clarify the physical picture of the symmetric phase.\\ 

\noindent
{\it Magnetic mass and crossover}

Consider the SU(2) Higgs model in three dimensions which is defined 
by the action
\beq\label{l3d}
S = \int d^3x \; \tr \left({1\over 2}W_{\mu\nu}W_{\mu\nu} + 
(D_{\mu}\PH)^\hc D_{\mu}\PH + \mu^2 \PH^\hc \PH 
+ 2 \lambda (\PH^\hc \PH)^2 \right) \, , 
\eeq
with 
\beq
\PH = {1\over 2} (\s + i \vec{\pi}\cdot \vec{\tau}) \, ,\quad 
D_{\mu}\PH = (\partial_{\mu} - i g W_{\mu})\PH\, ,\quad  
W_{\mu} = {1\over 2}\vec{\tau}\cdot \vec{W_{\mu}}\ .
\eeq
Here $\vec{W_{\mu}}$ is the vector field, $\s$ is the Higgs field, $\vec{\pi}$
is the Goldstone field and $\vec{\tau}$ is the triplet of Pauli matrices.
The gauge coupling $g$ and the scalar coupling $\lambda$ have mass dimension 
1/2 and 1, respectively. For perturbative calculations gauge fixing and 
ghost terms have to be added. The parameters of the three-dimensional Higgs 
model have been related to the parameters of the four-dimensional Higgs model 
at finite temperature by means of dimensional reduction \cite{jak}. 
In particular, variation of temperature corresponds to variation of the mass 
parameter $\m^2$.

We are interested in the propagators $G_\s$ and $G_W$ of Higgs field
and vector field, respectively, which at large separation $|x-y|$
fall off exponentially,
\bea\label{expo}
G_\s(x-y) &=& \left\langle \s(x)\s(y)\right\rangle \sim e^{-M |x-y|}\ ,\nn\\
G_W(x-y)_{\m\n} &=& \langle W_{\m}(x)W_{\n}(y)\rangle \sim e^{-m |x-y|}\ .
\eea 
For $\m \gg g^2$ one has $M \simeq \m$, whereas $m$ 
cannot be computed in perturbation theory.
A non-vanishing vector boson mass can be 
obtained from a coupled set of gap equations for Higgs boson and vector boson 
masses as follows. One shifts the Higgs field $\s$ around its vacuum 
expectation value $v$, $\s = v + \s'$, which yields the tree level masses
\beq
m_0^2 = {g^2\over 4}v^2\ ,\ M_0^2 = \mu^2 + 3\lambda v^2.
\eeq
The masses $m_0^2$ and $M_0^2$ are now expressed as
\beq\label{masses}
m_0^2 = m^2 - \delta m^2\ ,\ M_0^2 = M^2 - \delta M^2\ ,
\eeq
where $m$ and $M$ enter the propagators in the loop expansion, and
$\delta m^2$ and $\delta M^2$ are treated perturbatively as counter terms. 
Together with the mass resummation a vertex resummation is performed.
One then obtains a coupled set of gap equations for Higgs boson
and vector boson masses,
\bea\label{gaps}
\delta m^2 + \Pi_T(p^2 = -m^2, m, M, \xi) = 0\ ,\nn\\
\delta M^2 + \Sigma(p^2 = -M^2, m, M, \xi) = 0\ ,
\eea
where $\Pi_T(p^2)$ is the transverse part of the vacuum polarization tensor.
The calculation has been carried out in  $R_{\xi}$-gauge. In order to
obtain masses $M$ and $m$ which are independent of the gauge parameter
$\xi$, it is crucial to perform a vertex resummation in addition to
the mass resummation and to evaluate the self-energy terms on the mass shell
\cite{reb}. This yields the screening lengths defined in Eq.~(\ref{expo}).
`Magnetic masses' defined at zero momentum are gauge-dependent \cite{bfhw}.

Together with a third equation for the vacuum expectation value $v$, 
determined by the condition $\langle\s'\rangle = 0$, the gap equations
determine Higgs boson and vector boson masses for each set of values
$\m^2/g^4$ and $\lambda/g^2$. For negative $\m^2$ one finds a unique
solution, corresponding to the familiar Higgs phase, with masses close
to the results of perturbation theory. In the case of small
positive $\mu^2/g^4$ and sufficiently small $\lambda/g^2$ there exist
two solutions, corresponding to the Higgs phase and the symmetric phase
with a small, but finite vector boson mass, respectively. This is the 
metastability
range characteristic for a first-order phase transition. For large 
positive $\m^2/g^4$ only the solution corresponding to the symmetric
phase remains. Here the Higgs boson mass is $M \simeq \m$, and the
vector boson mass, which is rather independent of $\m$ and $\lambda$,
is given by
\bea\label{msm}
m_{SM} &=& C g^2\ ,\nn\\ 
C &=& {3\over 64\pi}(21 \ln 3 - 4) \simeq 0.28 \ .
\eea
Note that this value is rather close to the magnetic mass obtained from 
gap equations for the pure gauge theory \cite{nair,jackiw}
as well as to the three-dimensional confinement scale estimated in \cite{reu}.
However, the physical interpretation of the magnetic mass is controversial 
in the case of the pure gauge theory \cite{jackiw}. Furthermore, 
the magnetic mass (\ref{msm}) is consistent with the propagator mass 
obtained in a numerical lattice simulation in Landau gauge \cite{neu},
\beq
m_{SM}^{(L)} = 0.35 (1) g^2\, .
\eeq
The mass gap in the symmetric phase is a direct consequence of the non-abelian
gauge interactions. In the abelian Higgs model no vector boson mass is 
generated in the symmetric phase \cite{buphi2}.

\begin{figure}[t]

\vspace{-1.1cm}
 
\begin{center}
\leavevmode
\epsfysize=500pt
\epsffile{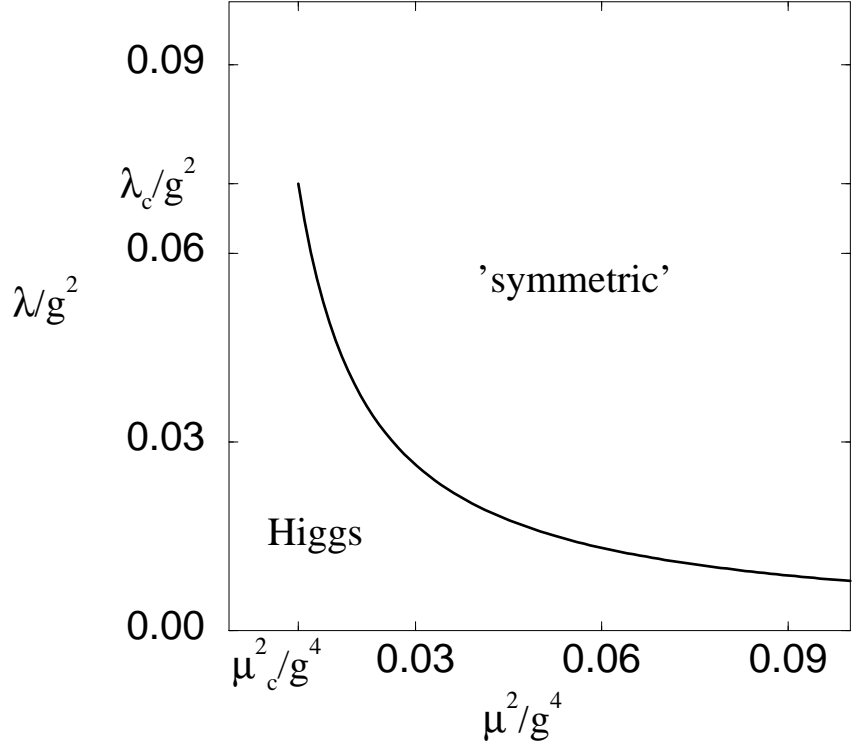}
 
\vspace{-6.0cm}

\end{center}
\caption[]{\label{phased}\it
Critical line of first-order phase transitions as given by Eq.~(\ref{crit}).}
 
\vspace*{0.5cm}
 
\end{figure} 

The size of the magnetic mass determines the critical Higgs boson mass where
the first-order transition changes to a crossover. Such a crossover behaviour 
was first observed for the four-dimensional finite-temperature Higgs model
in numerical lattice simulations for large Higgs masses by Evertz, Jers\'ak
and Kanaya, who also discussed in detail the phase diagram \cite{dam}. 
In connection with the average action approach the change to a crossover
was discussed in \cite{ber}.
A rough estimate of the critical Higgs mass, where a crossover behaviour
sets in, can be obtained as follows. Consider the one-loop effective 
potential in unitary gauge ($\vec{\pi} = 0$), 
\beq\label{v1l}
V_{1l} = {1\over 2} \m^2 \s^2 + {1\over 4}\lambda \s^4
         - {1\over 16\pi} g^3 \s^3\, ,
\eeq
where we have neglected the scalar contributions for simplicity.
At the beginning of the metastability range, $\mu^2=0$, the Higgs vacuum
expectation value is 
$\s_0 = 3g^3/(16\pi \lambda)$, which corresponds to
the vector boson mass
\beq
m_W (\m^2 = 0) = {3\over 32 \pi} {g^4\over \lambda}\, .
\eeq
It is reasonable to expect that the first-order phase transition disappears
at a critical scalar coupling where the vector boson mass in the Higgs phase 
reaches the magnetic mass of the symmetric phase. The condition 
$m_W (\m^2 = 0) = m_{SM}$
determines a critical coupling $\lambda_c$. The corresponding 
zero-temperature critical Higgs boson mass is given by 
\beq\label{crith}
\bar{m}^c_{H} = \left({3\over 4\pi C}\right)^{1/2}\bar{m}_W
              \simeq 74\ \mbox{GeV}\, ,
\eeq
where $\bar{m}_W$ is the zero-temperature vector boson mass. Eq.~(\ref{crith})
clearly shows that the crossover point is determined by the constant $C$,
i.e., the size of the magnetic mass. 
The obtained value of 
the critical Higgs mass agrees rather well with the result of recent 
numerical simulations \cite{lai}. In contrast, for vanishing magnetic mass the
first-order transition never changes to a crossover, while taking $C>1.0$ 
corresponding to the measured bound state mass $m_V$ (cf.~Eq.~(\ref{vec}), 
table~1) grossly underestimates $\bar{m}_H^c$.

From the effective potential (\ref{v1l}) one can easily determine the
critical line of the first-order phase transition. For 
$\hat{\lambda}(\m^2) < \lambda_c$ the conditions
\beq
0 = V_{1l}(\s_0)
  = \left.{\partial\over \partial \s} V_{1l}\right|_{\s=\s_0}
\eeq
yield for the critical line $\hat{\lambda}(\m^2)$,
\beq \label{crit}
{\hat{\lambda}(\m^2)\over g^2} = {1\over 128\pi^2} {g^4\over \mu^2}\ .
\eeq
The corresponding phase diagram is shown in Fig.~\ref{phased}. 
The critical value of 
the mass parameter at the crossover point is given by
\beq
{\m_c^2\over g^4} \simeq {C\over 8\pi}\, .
\eeq 
Using the matching relations to the finite-temperature Higgs model one
can evaluate the critical temperature as function of the zero-temperature
Higgs boson mass.\\

\noindent
{\it Gauge-invariant correlation functions}

It is expected that the SU(2) Higgs model has only a single phase, and
that the Higgs and the confinement regime are analytically connected
\cite{oster}. All physical properties of the model can be obtained by studying
correlation functions of gauge-invariant operators. This is of particular
importance for numerical lattice simulations where in general the gauge is 
not fixed.

In the literature the following operators for scalar states with 
$J^{PC}=0^{++}$ have been studied,
\bea
R(x) &=& \tr\left(\PH^{\hc}(x)\PH(x)\right)\, ,\\
L(x) &=& \tr\left((D_{\m}\PH)^{\hc}D_{\m}\PH(x)\right)\, ,\\
P(x) &=& {1\over 2} \tr\left(W_{\m\n}W_{\m\n}\right)
      = - {1\over 8g^2}\tr\left([D_{\m},D_{\n}][D_{\m},D_{\n}]\right)\, .
\eea
The standard operator for vector states with $J^{PC}=1^{--}$ is
\beq\label{vec}
V^a_{\mu}(x) = {1\over 2} 
\tr\left(\PH^{\hc}(x)\stackrel{\leftrightarrow}{D_{\m}}\PH(x)\tau^a\right)\, .
\eeq
In the numerical simulations \cite{mon,kajla,tep,neu,gue} screening masses 
have been determined from the 2-point functions of the operators $R$ and 
$V^a_{\m}$,
\bea
G_R(x-y) &=& \langle R(x) R(y) \rangle \sim  e^{-m_R |x-y|}\, ,\\
G_V(x-y)_{\m\n} &=& \langle V_{\m}(x) V_{\n}(y) \rangle \sim  e^{-m_V |x-y|}\, .
\eea
In \cite{tep} screening masses have also been measured for the
operators $L(x)$ and $P(x)$,
\bea
G_L(x-y) &=& \langle L(x) L(y)\rangle \sim e^{-m_L |x-y|}\, ,\\
G_P(x-y) &=& \langle P(x) P(y)\rangle \sim e^{-m_P |x-y|}\,.
\eea
The screening masses $m_R$, $m_V$, $m_L$ and $m_P$ have been determined
for positive and negative values of $\m^2$, i.e., in the symmetric phase
and in the Higgs phase. In the latter case, as expected, the results agree
with perturbation theory and gap equations. The value for $m_P$ is 
consistent with an intermedate state of two massive vector bosons $V$
contributing to $G_P$. This is the leading contribution if one expands
$P(x)$ in powers of $g^2$.

In the symmetric phase, however, the numerical results for $m_R$ and $m_V$
do not agree with the predictions of the gap equations. Since also in the
symmetric phase the vacuum expectation value of the Higgs field is
different from zero, it was suggested in \cite{buphi} that the magnetic
mass $m_{SM}$ should determine the asymptotic behaviour of $G_V$ 
(cf.~(\ref{fms})). The numerical simulations show no sign of this.

What is the connection between the gauge-dependent 2-point function
$G_W$ and the gauge-invariant 2-point function $G_V$? This question has
been addressed by Fr\"ohlich, Morchio and Strocci in their  
detailed study of the Higgs phenomenon in terms of gauge-invariant
operators \cite{froe}. As they have pointed out,  gauge-invariant
correlation functions are 
approximately proportional to gauge-dependent correlation
functions as calculated in standard perturbation theory, if for the
chosen gauge and renormalization scheme the fluctuations of the Higgs
field are small compared to the vacuum expectation value. For instance,
for the scalar correlation functions one has ($\s=v+\s', 
\langle \s'\rangle =0$),
\beq\label{fms}
\langle R(x) R(y)\rangle \sim v^2 \left(\langle \s'(x) \s'(y)\rangle
 + \mbox{\cal O}\left({\s' \over \langle \s \rangle}, {\vec{\pi}\over 
        \langle \s\rangle}\right)\right)\, .
\eeq
A measure for the relative size of the fluctuation terms is the ratio
\beq\label{ratio}
\zeta = {\langle \PH^{\hc}\PH \rangle \over \langle \s \rangle^2}\, .
\eeq
At one-loop order one obtains in $R_{\xi}$-gauge,
\bea\label{fluc}
\langle \PH^{\hc}\PH\rangle &=& \langle \s^2+\vec{\pi}^2\rangle
 = v^2 + \langle \s'^2 + \vec{\pi}^2\rangle \\
&=& v^2 - {1\over 4\pi}\left(M + 3 \sqrt{\xi} m \right)\, .
\eea
Here linear divergencies have been subtracted by means of dimensional
regularization. Deep in the Higgs phase, where $\m^2 < 0$, 
$v_0^2 = - \m^2/\lambda$,
$M^2_0 = 2\lambda v_0^2$ and $m_0^2 = g^2 v_0^2/4$, one finds
\beq
\zeta_H = 1 - {3\over 8\pi}\left(\sqrt{\xi} + 
   {2\over 3} {\sqrt{2\lambda}\over g}\right) {g\over v} + \ldots\, .
\eeq
In the relevant range of parameters one has $g/v < 1$. Hence, the deviation 
of $\zeta_H$ from 1 is small and ordinary perturbation theory is reliable. 

On the contrary, in the symmetric phase the situation is very different.
Here the gap equations yield for the vacuum expectation value $g/v
\simeq 10$. Inserting in the definition of the ratio (\ref{ratio}) solutions 
of the gap equations for $M$ and $m$ one finds in the symmetric phase
that $\zeta_{SM}$ deviates from 1 by more than $100\%$.
Hence, we cannot expect that the gauge-dependent 2-point functions give a 
good approximation to the gauge-invariant 2-point functions. 

\noindent
{\it Gauge-invariant screening masses}
                  
\begin{figure}[t]
\vspace{-5cm}
\begin{center}
\leavevmode
\epsfysize=200pt
\hspace*{-1.5cm}
\epsffile{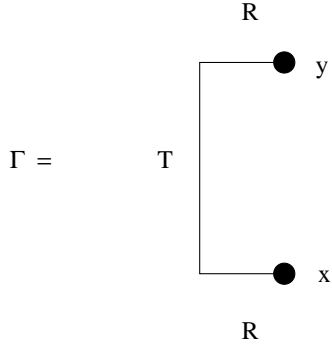}
\vspace{-0.5cm}
\end{center}
\caption[]{\label{gamma}\it The path $\Gamma$ for $G_{\PH}(T,R)$.}
\vspace*{0.5cm}
\end{figure}

What is the physical meaning of the propagator masses obtained from gap 
equations as well as numerical simulations in the symmetric phase? 
Several years ago the notion of a `screening energy' has been introduced
in connection with an analysis
of the SU(2) Higgs model at zero temperature \cite{kas}.
The authors considered the gauge-invariant correlation function
\beq \label{gphi}
G_{\PH}(T,R) = \left\langle\tr\left(\PH^{\hc}(y)U(\Gamma)\PH(x)
                     \right)\right\rangle\, ,
\eeq
where
\beq
U(\Gamma) = P \exp{\left(ig\int_{\Gamma}ds\cdot W\right)}
          \equiv U^{\hc}(R,y)U(T)U(R,x)\, ,
\eeq
and the path $\Gamma$
\beq
\Gamma \equiv \Gamma(y,R)\circ\Gamma(T)\circ\Gamma(R,x)
\eeq
is shown in Fig.~\ref{gamma}.
For large $T$, with $R$ fixed, an exponential fall-off was found,
\beq\label{mcon}
G_{\PH}(T,R) \sim e^{-m_{\PH}T}\, ,
\eeq
with $m_{\PH}$ being independent of $R$. In temporal gauge the 2-point
function takes the form,
\beq\label{ener}
G_{\PH}(T,R) = \left\langle \tr\left(\PH^{\hc}(x)U^{\hc}(R,x)e^{-HT}U(R,x)
               \PH(x)\right)\right\rangle\, ,
\eeq
where $H$ is the hamiltonian. Comparison of Eqs.~(\ref{mcon})
and (\ref{ener}) suggests that $m_{\PH}$ is the energy of a dynamical charge
bound by an external charge \cite{kas}. If the energy of the infinitely
heavy external charge is properly subtracted, $m_{\PH}$ corresponds to the
`constituent', or screening mass of the bound scalar $\PH$. In the case $R=0$ 
the 2-point function $G_{\PH}$ reduces to the gauge-invariant propagator
\beq
\hat{G}_{\PH}(x-y)=\left\langle \tr\left(\PH^{\hc}(y)U_{yx}\PH(x)\right)
                   \right\rangle
                   \sim e^{-m_{\PH}|x-y|}\, ,
\eeq
where $U_{yx}$ is the non-abelian phase factor along the straight line
from $x$ to $y$. 

The definition of a screening mass for the vector boson is completely
analogous. The obvious definiton is
\beq
\hat{G}_W(x-y)_{\m\n\r\s} = \left\langle W^T_{\m\n}(y)
U^{\cal A}_{yx} W_{\r\s}(x)\right\rangle
\sim e^{-m_W |x-y|}\, .
\eeq
where the superscript ${\cal A}$ denotes SU(2) matrices in the adjoint
representation.

The contribution from the phase factor to the masses $m_{\PH}$ and
$m_W$, which depend on the mass parameter $\m^2$, are linearly divergent. 
Renormalized screening masses can be 
defined by matching $m_{\PH}$ and $m_W$ to the masses $m_R$ and $m_V$
of the gauge-invariant correlation function at some value $\m^2_0$ in the 
Higgs phase. The corresponding screening masses $m_{\PH}(\m^2;\m^2_0)$ and
$\m^2_W(\m^2;\m^2_0)$ satisfy the boundary conditions
\beq
m_{\PH}(\m^2_0;\m^2_0) = m_R(\m^2_0)\quad ,\quad
m_W(\m^2_0;\m^2_0) = m_V(\m^2_0)\; .
\eeq
These screening masses, as functions of $\m^2$, should essentially coincide
with the solutions $M(\m^2)$ and $m(\m^2)$ of the gap equations, respectively.

\begin{figure}[t]

\begin{center}
\leavevmode
\epsfysize=250pt
\hspace*{1.5cm}
\epsfbox[20 400 620 730]{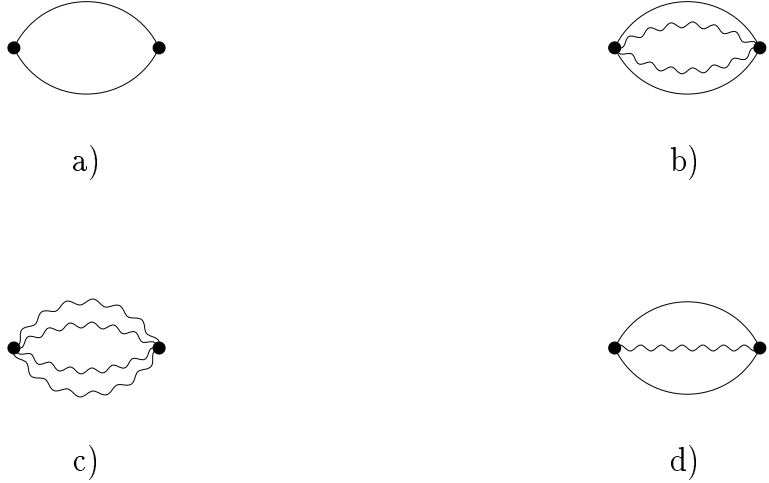}
 
\vspace{-1.5cm}

\end{center}
\caption[]{\label{2pt}\it
  Two-point functions to leading order for a constituent model.}
\vspace*{0.5cm}
\end{figure}
What is the role of the screening masses in the correlation functions
of gauge-invariant operators? As discussed above, the fluctuations dominate
in the symmetric phase. Hence, one may expect that multi-particle states
of `constituent' scalar and vector bosons dominate the exponential fall-off
of the 2-point functions. This is in the spirit of 
previously proposed bound state models \cite{hel,dosch}.
According to Fig.~\ref{2pt},
for $G_R$ this should be a $(\PH^{\hc}\PH)$ state (a),
for $G_L$ a $(\PH^{\hc}W W \PH)$ state (b), for $G_P$ a $(W W W W)$ state (c) 
and for $G_V$ a $(\PH^{\hc}W\PH)$ state (d). 
Here we have identified a covariant
derivative $D_{\m}$ with a constituent vector boson $W$, since for bound
states in the symmetric phase an expansion in powers of $g^2$ is not
justified. 
Neglecting binding effects, this yields the mass formulae
\bea
m_R \simeq 2 m_{\PH}\,\quad m_L \simeq 2 m_{\PH} + 2m_{W}\, 
\quad m_P \simeq 4 m_W\, \quad m_V \simeq 2 m_{\PH} + m_W\, .
\eea 
These relations can be compared with results from lattice simulations.
A screening mass $m_W$ for the vector boson was determined in \cite{neu},
$m_W = 0.35(1) g^2$. No scalar screening mass has been measured so far,
hence we choose $m_{\PH} = m_R/2$. This yields three predictions for
$m_L$, $m_P$ and $m_V$ which are compared with the results of \cite{tep}
in table~1. The 
qualitative agreement supports the constituent picture.

\renewcommand{\arraystretch}{1.5}
\begin{table}[h]
\begin{center}
\begin{tabular}{|c|cccc|c|}
\hline
    &&$J^{PC}=0^{++}$ & & &$J^{PC}=1^{--}$    \\
    &$ R $ &$ L $ & &$ P $ &$ V $  \\ \hline\hline
lattice simulations \cite{tep}
& 0.839(15) & 1.47(4) & & 1.60(4) & 1.27(6) \\ \hline
constituent model & -  & 1.54$\;$ & & 1.40$\;$ & 1.18$\;$  \\ \hline
\end{tabular}
\end{center}
\caption[]{\it Comparison of screening masses from lattice simulations
and a constituent model. $m_R$ is used to fix the constituent scalar mass.}
\end{table}

The proposed picture can be tested by measuring
the gauge-invariant propagators $\hat{G}_{\PH}$ and $\hat{G}_W$ as
functions of $\m^2$. The masses $m_{\PH}(\m^2;\m^2_0)$ and 
$m_W(\m^2;\m^2_0)$ should 
behave like the solutions $M(\m^2)$ and $m(\m^2)$ of the gap equations.
In particular, at a first-order transition from the Higgs phase to the
symmetric phase, both screening masses should jump to smaller values.
With increasing $\lambda/g^2$ the jump should decrease and eventually
vanish at the critical coupling where the crossover behaviour sets in.\\

\noindent
{\bf Acknowledgements}\\
We would like to thank K.~Jansen, M.~Laine, M.~L\"uscher, I.~Montvay and 
M.~Teper for helpful discussions.\\

\end{document}